%% file: main.tex
\documentclass[conference]{IEEEtran}
\usepackage{cite}
\usepackage{amsmath,amssymb,amsfonts}
\usepackage{graphicx}
\usepackage{textcomp}
\usepackage{xcolor}
\usepackage{comment}

\usepackage{mathptmx} 
\usepackage{fancyhdr}
\usepackage[normalem]{ulem}
\usepackage[hyphens]{url}
\usepackage[final]{microtype}
\usepackage[keeplastbox]{flushend}
\usepackage{adjustbox}
\usepackage{multirow}
\usepackage[bookmarks=true,breaklinks=true,letterpaper=true,colorlinks,linkcolor=black,citecolor=blue,urlcolor=black]{hyperref}
\usepackage{enumitem}
\usepackage[mathscr]{eucal}
\usepackage{algorithm}  
\usepackage{algorithmicx}
\usepackage{algpseudocode}  
\usepackage{todonotes}
 
\IEEEoverridecommandlockouts
\title{CAMA: Energy and Memory Efficient Automata Processing in Content-Addressable Memories} 
\author{\IEEEauthorblockN{Yi Huang\IEEEauthorrefmark{1}, Zhiyu Chen\IEEEauthorrefmark{1}, Dai Li and
Kaiyuan Yang}
\IEEEauthorblockA{Department of Electrical and Computer Engineering, Rice University, Houston TX}
}

\begin{document}
\maketitle
\def\thefootnote{*}\footnotetext{These authors contributed equally to this work}\def\thefootnote{\arabic{footnote}}
\thispagestyle{plain}
\pagestyle{plain}


\begin{abstract}

    Accelerating finite automata processing is critical for advancing real-time analytic in pattern matching, data mining, bioinformatics, intrusion detection, and machine learning. Recent in-memory automata accelerators leveraging SRAMs and DRAMs have shown exciting improvements over conventional digital designs. However, the bit-vector representation of state transitions used by all state-of-the-art (SOTA) designs is only optimal in processing worst-case completely random patterns, while a significant amount of memory and energy is wasted in running most real-world benchmarks. 
    


    We present CAMA, a Content-Addressable Memory (CAM) enabled Automata accelerator for processing homogeneous non-deterministic finite automata (NFA). A radically different state representation scheme, along with co-designed novel circuits and data encoding schemes, greatly reduces energy, memory, and chip area for most realistic NFAs. CAMA is holistically optimized with the following major contributions: (1) a 16$\times$256 8-transistor (8T) CAM array for state matching, replacing the 256$\times$256 6T SRAM array or two 16$\times$256 6T SRAM banks in state-of-the-art (SOTA) designs; (2) a novel encoding scheme that enables content searching within 8T SRAMs and adapts to different applications; (3) a reconfigurable and scalable architecture that improves efficiency on all tested benchmarks, without losing support for any NFA that's compatible with SOTA designs; (4) an optimization framework that automates the choice of encoding schemes and maps a given NFA to the proposed hardware. 
    

    Two versions of CAMA, one optimized for energy (CAMA-E) and the other for throughput (CAMA-T), are comprehensively evaluated in a 28nm CMOS process, and across 21 real-world and synthetic benchmarks. 
    CAMA-E achieves 2.1$\times$, 2.8$\times$, and 2.04$\times$ lower energy than CA, 2-stride Impala, and eAP. CAMA-T shows 2.68$\times$, 3.87$\times$ and 2.62$\times$ higher average compute density than 2-stride Impala, CA, and eAP. Both versions reduce the chip area required for the largest tested benchmark by 2.48$\times$ over CA, 1.91$\times$ over 2-stride Impala, and 1.78$\times$ over eAP.

\end{abstract}

\input{S1.tex}
\input{S2.tex}
\input{S3.tex}

\input{S4.tex}

\input{S5.tex}
\input{S6.tex}
\input{S7.tex}


\bibliographystyle{IEEEtranS}
\bibliography{refs}


\end{document}

%% file: S1.tex
\section{Introduction}

Real-time stream processing plays an increasingly important role in the era of big data and the internet of things. Non-deterministic Finite Automaton (NFA) is a popular computation model to perform complex pattern matching described by specification languages like regular expressions. NFA has shown promising applications in network security~\cite{li_cam_2020,pao_string_2011,tran_trung_hieu_memory_2013,yu_fast_network_2006}, data mining~\cite{wang_pattern_mining_2016,saredni_subtree_2017}, natural language processing~\cite{sadredini_scalable_2018}, bioinformatics~\cite{bo_grna_2018,roy_discovering_2016}, finance~\cite{aho_bibl_1975}, machine learning~\cite{bo_entity_2016}, and runtime verification of cyber-physical and embedded systems~\cite{alur_model_2001}. These applications demand high-speed processing of large and complex NFAs on real-time data collected from sensors, networks, and various system traces. In addition, high energy efficiency and memory efficiency (defined as the size of NFA that can be stored in a given amount of memory or chip footprint) are always highly desired in both cloud and embedded applications. 


It is well known that finite automata processing, like many other data- and memory-intensive computing problems, requires frequent, yet irregular and unpredictable memory access on general-purpose processors, leading to limited throughput and high power consumption on modern CPU and GPGPU architectures~\cite{wadden_anmlzoo_2016,lenjani_cache_miss_2014,liu_efficient_2011-1}. FPGA solutions offer much higher speed, but are often bottlenecked by the severe routing congestions when dealing with a large number of complex patterns\cite{rahimi_grapefruit_2020,xie_reapr_2017}. Even in application-specific digital accelerators, the memory access challenges limit the total number of states and transitions that can be processed in parallel, ultimately limiting their throughput and efficiency~\cite{tandon_hawk_2016,lunteren_designing_2012}. 



To circumvent these challenges, in-memory architectures are increasingly being explored. Directly processing the NFA states and transitions inside memories enjoys huge memory access and computing parallelism. It also lowers energy by combining memory access with logic operations. 
For instance, 
the seminal Automata Processor (AP) design~\cite{dlugosch_efficient_2014,wang_apview_2016} outperform x86 CPUs by 256×, GPGPUs by 32×, and accelerator XeonPhi by 62× in the ANMLZoo benchmark suite\cite{wadden_anmlzoo_2016,subramaniyan_cache_2017}, while Cache Automaton (CA)~\cite{subramaniyan_cache_2017} achieves 3.9× speedup over HARE\cite{gogte_hare_2016} and 3× speedup over UAP\cite{fang_fast_2015}.



Existing in-memory automata processors all employ the bit vector representation of the labels of states, i.e. the symbol classes accepted by each state, enabling \emph{state matching} in standard RAMs. 
However, across real-world and synthesized benchmarks~\cite{becchi_workload_2008,wadden_anmlzoo_2016}, it is found that only 3$\%$ of total processing resources are utilized in 86$\%$ of the time on average~\cite{sadredini_impala_2020}.
Thus, in most scenarios, the 256-bit vector representation for state matching is inefficient in memory usage, which also leads to performance and energy penalties due to the larger physical dimension of memories. E. Sadredini et al.~\cite{sadredini_impala_2020} proposed a high-throughput SRAM-based design, Impala, to tackle this particular issue by using two 16$\times$256 6T SRAM banks to function like one 256$\times$256 6T SRAM in CA. However, this scheme consumes more energy because of the doubled memory peripheral, which is the dominant source of energy.



In this work, we take an orthogonal approach and propose the first \textbf{C}ontent \textbf{A}ddressable \textbf{M}emory (CAM) enabled \textbf{A}utomata accelerator, namely \textbf{CAMA}, whose memory and energy usage scale down with the average symbol class size of a given NFA, without any performance overhead. CAM is naturally suitable for fixed-width bit string matching~\cite{pagiamtzis_content-addressable_2006}, which has been extensively used in networking for IP lookup and packet classification. Because CAMs are built purely with CMOS devices, it also enjoys perfect scalability to cutting-edge CMOS process nodes~\cite{yabuuchi_12-nm_2018}. However, existing CAM-based hardware for complex pattern matching mostly implement Deterministic Finite Automaton (DFAs)~\cite{peng_chain-based_2011-1} constructed from either AC algorithm or regular expressions \cite{bremler-barr_compactdfa_2014,alicherry_high_2006,yun_efficient_2012}. They cannot handle many complex applications beyond packet inspection, because of the well-known state explosion issues when converting an expressive NFA to a DFA~\cite{dlugosch_efficient_2014,wang_apview_2016}. To the best of our knowledge, CAMA is the first CAM-based hardware targeting generic NFA processing. 

In CAMA, CAMs perform \emph{state matching} by directly comparing the input symbol with the symbol classes of all enabled states, instead of converting the input symbol to a bit vector representation for comparison like in RAMs with single-row access. This change apparently results in a much narrower memory bit-width, but considering CAM's larger footprint and access energy than SRAMs, the remaining design challenges are to reduce the average number of CAM entries for each state and the overall hardware footprint and energy. Thus, we propose tightly coupled optimizations across circuits, architecture, and algorithms to (1) re-purpose standard 8-transistor (8T) SRAMs for \emph{state matching}, which is made possible by a specially crafted data encoding scheme, (2) adapt the CAM bit-width to NFAs with different sparsity, and (3) compresses multiple symbols into a single CAM entry. 
With these co-designs, CAMA increases the total number of CAM entries 
by merely 13\% but reduces the overall state matching memory area by 3.6$\times$ over CA, a SOTA in-SRAM automata accelerator. 


On the other hand, \emph{state transition} is another critical step to optimize. A. Subramaniyan et al.~\cite{subramaniyan_cache_2017} proposed the use of 8T SRAMs as fully connected crossbar interconnects, overcoming the limited flexibility of hardwired interconnect in AP. 8T SRAMs support logic OR operations among multiple activated rows, without destructing stored data. Later, E. Sadredini et al.~\cite{sadredini_eap_2019} observed that most applications have sparse state transitions in the shape of a diagonal matrix, and propose the Reduced CrossBar (RCB) to save area. Inspired by RCB, CAMA employs a 128$\times$128 compact local switch compatible with the CAM-based \emph{state matching}. The proposed local switch has a simpler and smaller physical implementation than that in eAP~\cite{sadredini_eap_2019}, and supports a novel reconfiguration scheme to be converted into a full crossbar (FCB) mode, which provides sufficient routing resources for any NFA. 



In summary, this paper makes the following contributions:
\begin{itemize}[noitemsep]
\item We propose CAMA, the first CAM-enabled, low-area, and energy-efficient automata processor for homogenous NFA in streaming processing, which replaces the bit vector representation in SOTA RAM-based designs by direct symbol matching in CAMs. CAMA is holistically optimized across circuits, architecture, and algorithms, in order to improve energy, area, and compute density while maintaining similar programmability as prior arts.

\item We re-purpose 8T SRAM arrays to behave like 16T Ternary CAMs and perform \emph{state matching} on programmable symbol classes at greatly reduced area and energy costs. This is made possible by the proposed data encoding and symbol class compression methods. All CAM entries will be encoded when storing into the memory, while the encoding of inputs will be performed online with a 256$\times$32 SRAM-based encoder. The encoder only occupies less than 0.11$\%$ energy on average. 

\item We develop a reconfigurable local crossbar switch with 8T SRAMs for \emph{state transition} in CAMA, which features a simpler and smaller physical implementation than prior arts. It can be configured as an RCB for most applications or a full crossbar (FCB) for dense NFAs, in order to maintain generalizability to any NFA. 


\item We create an optimization framework that automatically analyzes the homogenous NFA in an MNRL/ANML file, and chooses the optimal encoding scheme, the code length, and the CAMA operation mode. The toolchain also maps the optimized NFA to the hardware.

\item Two versions of CAMA designs in the 28nm CMOS process, one optimized for energy (CAMA-E) and the other for throughput (CAMA-T), are rigorously evaluated across 21 common real-world and synthetic benchmarks.
Compared to SOTA designs, CAMA-E achieves $> 2\times$ lower energy, while CAMA-T shows $> 2.6\times$ higher compute density (i.e. throughput per unit area). Both CAMA versions reduce the area required for the largest benchmark by 1.91$\times$, 1.78$\times$ and 2.48$\times$ over 2-stride Impala, eAP, and CA.

\end{itemize}

%% file: S2.tex
\section{Background}

\subsection{Non-Deterministic Finite Automata (NFA)}
A Non-deterministic Finite Automaton (NFA) is defined as a 5-tuple (Q,$\Sigma$,$\delta$,$q_{0}$,F), in which the Q represents the set of states, $\Sigma$ is a set of symbols, also called the input symbol alphabet, $q_{0}$ means the set of start states, F represents the set of reporting states and $\delta$ is defined as the transition function. The transition function $\delta$(Q, $\alpha$) gives out the set of states which are the next states starting from Q on input $\alpha$. Compared to deterministic FA (DFA), the NFA holds multiple active states at the same time and supports multiple transitions on a single input. 
\par
\begin{figure}[t]
      \centering
      \vskip -0ex
      \includegraphics[width=\columnwidth]{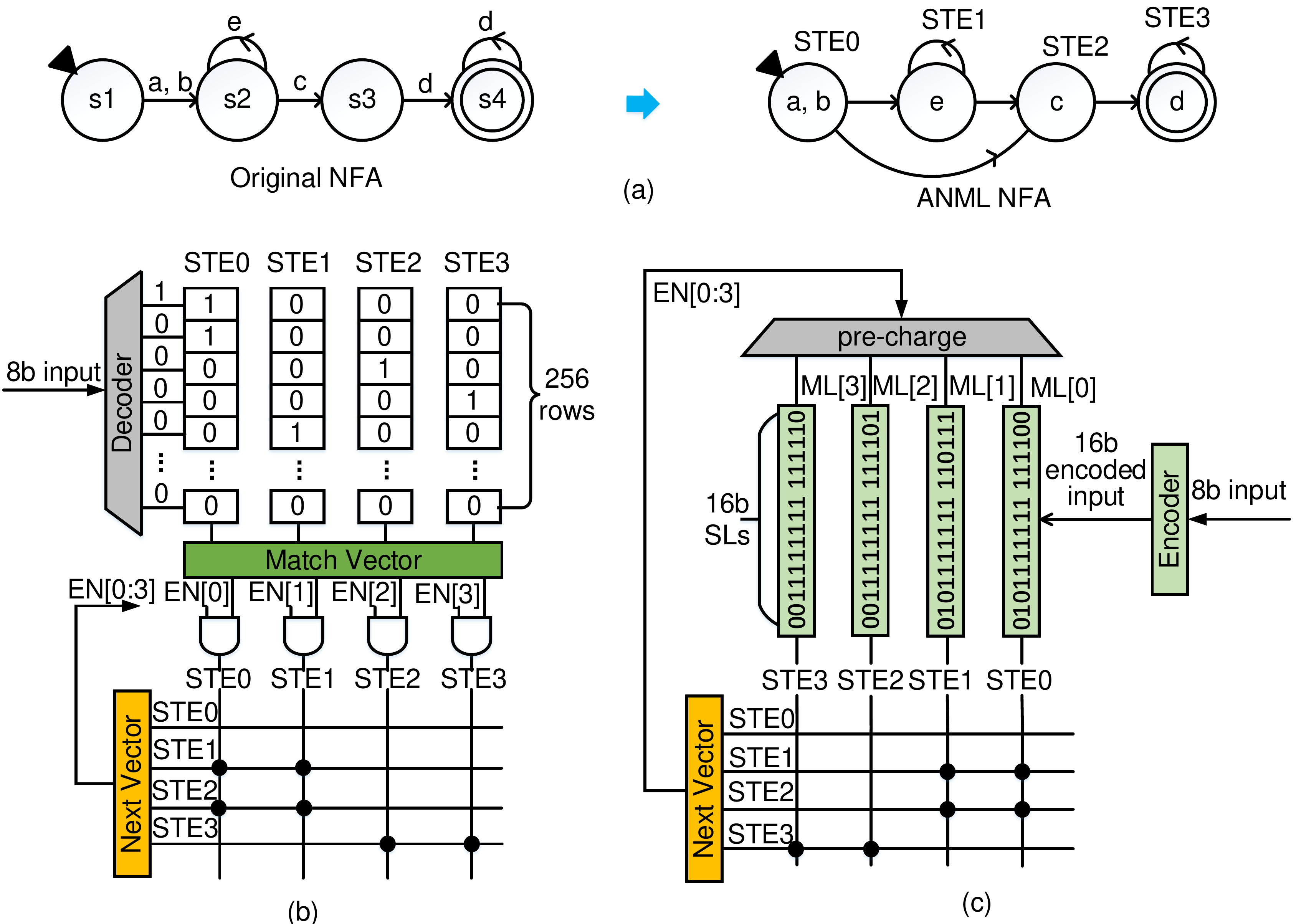}
      \vskip -2.5ex
      \caption{(a) Classical NFA and ANML-NFA; and principles of (b) CA, a representative in-memory automata accelerator, and (c) the proposed CAMA using Two-Zeros prefix encoding.}
      \label{NFA}
      \vskip -2.5ex
  \end{figure}

We adopt the ANML-NFA\cite{wadden_anmlzoo_2016,glushkov_abstract_1961}, or homogenous NFA, as our automaton model. A homogenous NFA is hardware-friendly and used by all SOTA designs~\cite{subramaniyan_cache_2017,dlugosch_efficient_2014,sadredini_impala_2020,sadredini_eap_2019}. 
Algorithms to convert a classical NFA into a hardware-friendly ANML-NFA have been well studied. Figure~\ref{NFA}(a) shows the classical NFA and the ANML-NFA for regular expression (a$\vert $b)$e^{*}$c$d^{+}$. The states in the ANML-NFA are often called state transition elements (STEs). Here the NFA alphabet is \{a, b, c, d, e\} and the states or STEs with a triangle on them are start states. Any state can be 
converted to a start state by enabling it on every input. States with double circles denote reporting/closing states in NFAs. 

\subsection{In-Memory Automata Accelerator}
In-memory automata accelerators are promising hardware solutions for general-purpose high-speed and high-efficiency automata processing, by circumventing the ``memory walls". Figure~\ref{NFA}(b) shows the working principle of CA~\cite{subramaniyan_cache_2017}, a representative design in SRAM. 

All existing in-memory automata processors execute an automaton in two steps: \emph{state matching} and \emph{state transition}. In \emph{state matching}, the static/dynamic RAM stores STEs in columns, with bit vector representations of each STE's symbol class. Every streaming input symbol is decoded to a one-hot address and activates one row in the state matching table (the RAM). Then the columns (STEs) comprising the input symbol, i.e. those storing `1's in the corresponding rows, will read out `1's, indicating potentially active STEs in the \emph{Match Vector}. Next, in \emph{state transition} step, \emph{Match Vector} is first logically ANDed with last cycle’s transition list (\emph{Next Vector}), to obtain the current active states. If any active state is also a reporting state, the result will be returned. In the meantime, an updated \emph{Next Vector} is generated by passing the active state vectors through a pre-programmed routing network. Figure~\ref{NFA}(b) illustrates how the regex (a$\vert$b)$e^{*}$c$d^{+}$ is mapped to CA\cite{subramaniyan_cache_2017}. Since this regex requires four STEs, only four memory columns are utilized in the 256$\times$256 state matching SRAM bank. A pre-programmed 4$\times$4 routing network (connections indicated by dots in Figure~\ref{NFA}b), which is a small subset of the actual 256$\times$256 local switches implemented in 8T SRAM, updates the transition list (\emph{Next Vector}) based on the current active STEs. As an example, assume STE0 is a current active STE that has a connection with STE1 and STE2. The next input symbol is `e' that activates Row5. The ANDed result on STE0's transition (EN[3:0]) and the \emph{Match Vector} determines STE1 as a current active STE for the next cycle.

\par
\begin{figure}[t]
      \centering
      \includegraphics[width=0.8\columnwidth]{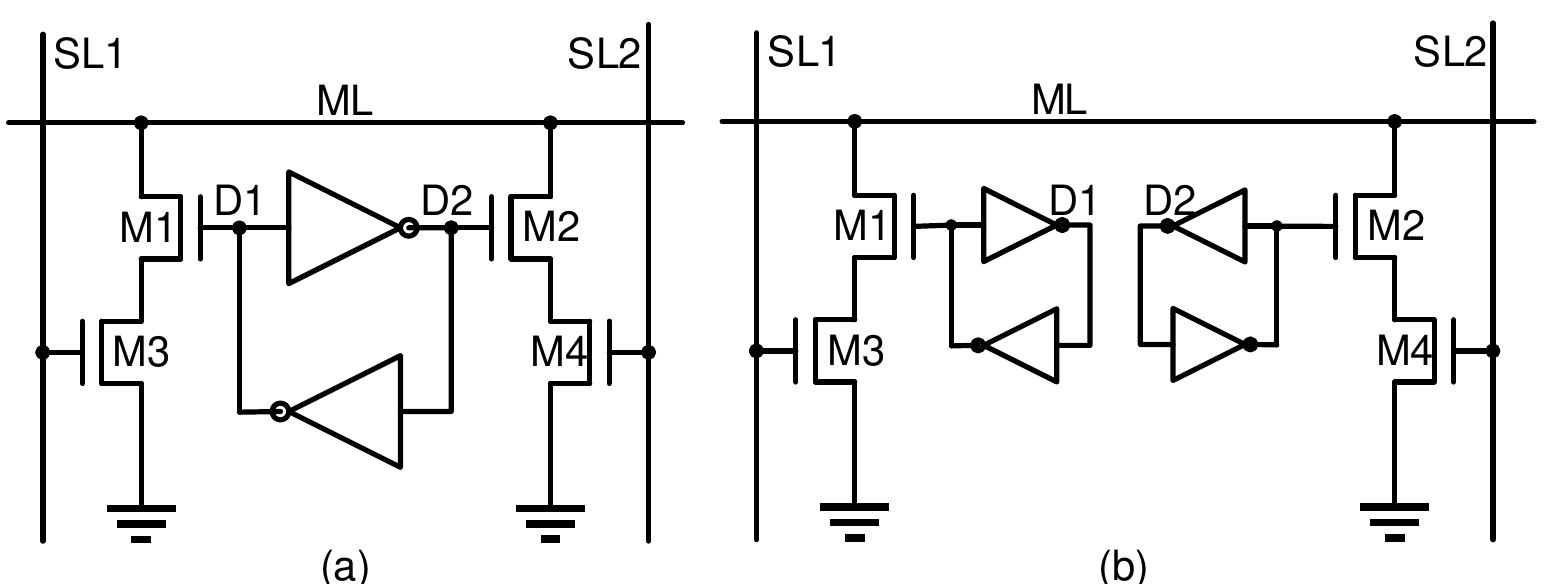}
      \vskip -2.5ex
      \caption{Circuit design for conventional (a)BCAMs and (b)TCAMs.}
      \label{cell}
      \vskip -2.5ex
  \end{figure}
\subsection{Content-Addressable Memories (CAM)}

\noindent\textbf{Binary CAM (BCAM)~\cite{pagiamtzis_content-addressable_2006}:}
Figure~\ref{cell}(a) shows the design of a standard 10-transistor (10T) NOR-type BCAM. The two pull-down paths from the match line (ML) to ground, formed by M1 through M4, implement the logic XNOR between the input search data on the differential search lines (SL1/SL2) and the differential stored data (D1/D2) in the 6T SRAM cell. Notice that bitlines (BLs), wordlines (WLs), and two access transistors for normal SRAM read/write are omitted for simplicity. A mismatch between D1/D2 and SL1/SL2 will pull down the precharged ML through one of the NMOS pairs, while a match will disconnect both pull-down paths. When multiple cells are connected to a single ML to form a CAM word, all pull-down paths effectively perform the logic NOR operation. ML keeps the precharged state (i.e. the matching condition) only if all cells are in the matching state.


\par

\noindent\textbf{Ternary CAM (TCAM)~\cite{pagiamtzis_content-addressable_2006}:}
Unlike the BCAMs that only store and search logic `1' and `0', TCAM cells support a new logic state `X' (i.e. \textit{don't care}) which matches both logic `1' and logic `0', and enables the so-called wildcard search. Figure~\ref{cell}(b) shows the implementation of the standard 16T NOR-type TCAM cell. Two SRAM bit cells are required to encode the 3 logic states, where `01' represents logic `0', `10' represents logic `1', and `11' is logic `X'. As a result, `X' will always disconnect the two pull-down paths and indicate a match regardless of the input data, while searching operations on `0' and `1' are identical to that of BCAM. In addition to storing `X', TCAM also allows searching `X' by setting both SL1 and SL2 to `0'. Similarly, this will result in a ``match" output due to the disconnected pull-down NMOS (M3/M4).


%% file: S3.tex
\section{Principles of CAM-Based Automata (CAMA)}
In this section, we first discuss the limitation of prior works and then summarize how CAMA tackles the challenges. 

\subsection{Limitations of prior works}

\textbf{Limitations in State Matching.} Existing in-memory automata accelerators are all based on bit vector (one-hot encoding) representation for state matching, which has a constant cost for each state regardless of the actual transition symbols. This leads to inefficient memory usage and high energy in most applications/NFAs, where each state only matches a small number of symbols. For instance, more than 86$\%$ of the states are matched against at most eight symbols across 21 benchmarks in~\cite{sadredini_impala_2020}, which means only 3$\%$ ($\approx 8/256$) of the cells in memory columns are utilized. Besides, the small size of the alphabet in some applications also causes underutilization of the state matching resources in CA, e.g., ExactMath~\cite{becchi_workload_2008} has only 114 symbols so that more than half of the SRAM rows remain unused.
Impala~\cite{sadredini_impala_2020} made the first attempt to tackle this issue by splitting 8-bit symbols to 4-bit symbols so that 256-bit one-hot state matching SRAMs can be replaced by two 16$\times$256 ones. However, because SRAM periphery occupies $\geq$80\% of SRAM access energy, reading two 16$\times$256 SRAM banks consumes more energy than reading a single 256$\times$256 bank. As a result, the energy efficiency of Impala turns out to be lower than CA~\cite{subramaniyan_cache_2017} that uses conventional 256-bit one-hot encoding. 
\par
\textbf{Limitations in State Transition.} Conventionally, in-memory automata architectures, such as CA and Impala, implements state transition by a full crossbar (FCB) local switch built with 8T SRAMs. The local switches connect each STE to any other STEs in the same state matching bank including itself and thus the area overhead is proportional to $N\times N$, where $N$ is the STE numbers in each state matching bank ($N=256$ in CA\cite{subramaniyan_cache_2017} and Impala\cite{sadredini_impala_2020}). 
However, in most real-world applications, the NFA transitions can be clustered into connected components (CCs) with no transitions between different CCs. Under the breadth-first search (BFS) mapping in eAP\cite{sadredini_eap_2019}, the state transition has a diagonal connectivity pattern in most applications, which means most valid transitions are located near the diagonal of the local switches. Therefore, only a small fraction of the SRAM cells are utilized. \cite{sadredini_eap_2019} shows that the average local switch utilization is merely 0.48$\%$ across 19 benchmarks in ANMLZoo\cite{wadden_anmlzoo_2016} and Regex\cite{becchi_workload_2008}.


\subsection{State Matching in CAMA}

Since the memory under-utilization comes from the overlong one-hot address, CAM is a promising alternative to reduce the code length of symbols, without doubling the periphery energy like in Impala\cite{sadredini_impala_2020}. In a CAM-based state matching design, the ANML-NFA states are no longer stored in the SRAM columns and input symbols will not be decoded into one-hot addresses to access SRAM rows. Instead, NFA states are encoded based on their symbol classes and mapped into CAM entries (CAM states). Here, the input symbols are directly sent to SLs in the CAM without the need of address decoders. Ideally, this modification reduces the bit-width of matching tables from $\mathscr{A}$ to $log_2\mathscr{A}$ bits ($\mathscr{A}$ is the size of the alphabet) while only requires one readout periphery. Furthermore, as Figure~\ref{NFA}(c) shows, the AND operation in the conventional architecture can be replaced by only precharging the columns enabled by transitions in the CAM through controllable column prechargers. This selective enabling scheme will greatly reduce CAM energy and is adopted in the proposed energy-optimized CAMA-E. 

A na\"ive implementation of this scheme is to use Binary CAM (BCAM) and the ASCII encoding, which only supports exact string match. It requires the number of CAM entries the same as the total symbol class size for all states. For example, RandomForest in ANMLZoo has a larger than 100 average symbol class size. Implementing it with a BCAM implementation will need $>$ 100 entries for each state, diminishing all the expected benefits from using CAM. Considering the cell-level larger energy and area of CAMs over SRAMs, this na\"ive design will only be advantageous when the symbol class size is very small (e.g. $\leq$ 3). 

Thus, we will need CAMs capable of storing and matching several symbols within a single entry, which is called compressed symbol class. To perform such symbol searches, the \emph{state matching} CAM needs to operate like a Ternary CAM (TCAM), where the \emph{don't care} state `X' allows compression of multiple symbols into a single entry. However, a conventional 16-transistor TCAM cell is about three times larger than 6T SRAM cells and consumes more energy, limiting the area and energy benefits of the proposed architecture. To alleviate these limitations and boost the CAMA efficiency, we propose holistic optimizations in both circuit and algorithm levels which will be discussed in \S \uppercase\expandafter{\romannumeral4} and \S \uppercase\expandafter{\romannumeral5}.

\subsection{State Transition in CAMA}

We propose a reconfigurable reduced crossbar (RRCB) compatible with the CAM-based \emph{state matching}, to achieve a dense switch for \emph{state transition}, inspired by the RCB approach in eAP~\cite{sadredini_eap_2019}. The RCB reduces the crossbar size from $256\times 256$ in CA~\cite{subramaniyan_cache_2017} to $96\times 96$, by remapping the sparsely and diagonally located transitions in FCB into a denser storage in RCB. To support NFAs where the transitions are too dense to fit into a single RCB (e.g., transitions in EntityResolution~\cite{wadden_anmlzoo_2016}), it reuses state-matching SRAMs for FCB state transitions. However, in the proposed CAM-based architecture, CAM banks are too small to function as FCBs. Additionally, the physical implementation of RCB in eAP requires three bit line (BL) segments in each column and three word lines (WLs) in each row, leading to congested routing in practical SRAM layout with potentially much larger memory footprint and larger energy. In the proposed 128$\times$128 RRCB for CAMA, transitions in the 8T SRAM crossbar are evenly allocated for most applications with a different mapping strategy than RCB (details in \S IV). The RRCB requires only two segments of WLs and BLs, which can be implemented within standard 8T SRAMs without any area overhead. It can further be reconfigured into a 128$\times$128 FCB by altering the connection of WLs and BLs for dense transitions. With the new reconfiguration scheme, reusing the state matching CAM as FCB is no longer needed. 


\subsection{Working Example of CAMA}

Figure~\ref{NFA}(c) shows an working example of the CAMA on a regular expression (a$\vert $b)$e^{*}$c$d^{+}$. In this example, the ASCII alphabet can be encoded to 16 bits and fit into 4 columns in CAMA's 16$\times$256 state matching bank with no state increase over bit vector representation. Unlike SOTA designs that encode 8-bit inputs into 256-bit row addresses in Figure~\ref{NFA}(b), CAMA's 8-bit input will be encoded into 16-bit input and directly sent to the search line. Ideally, only 8-bit input code is required, but we utilize 16-bit encoding to balance the CAM entries and the code length (which are described in greater detail in \S \uppercase\expandafter{\romannumeral4} and \S \uppercase\expandafter{\romannumeral5}). As for the state transition, this example is using an FCB for simplicity, but the actual implementation of CAMA is based on the proposed 128$\times$128 RRCB design as discussed in \S \uppercase\expandafter{\romannumeral4}. 


%% file: S4.tex
\section{Circuit Design and Optimization}
\begin{figure}[t]
      \centering
      \includegraphics[scale=0.6]{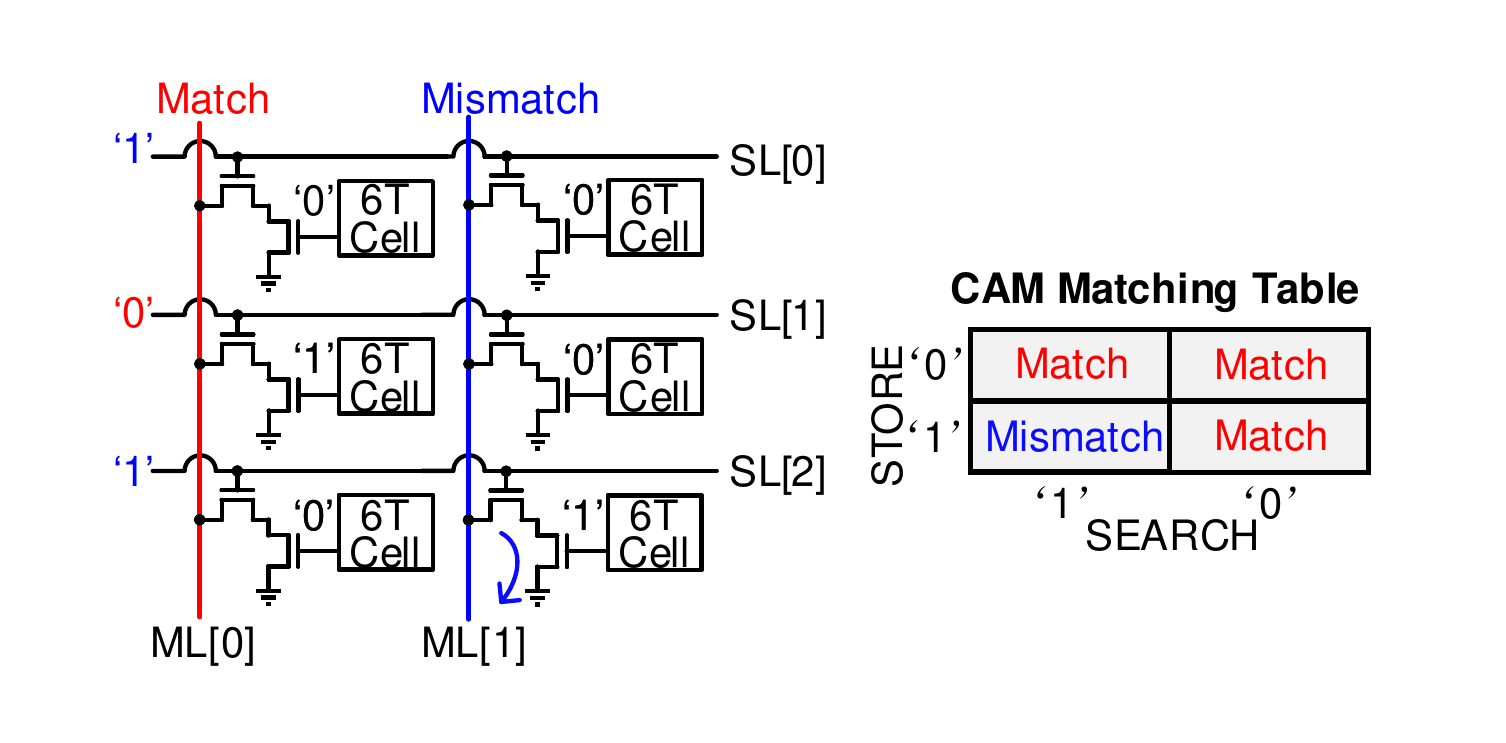}
      \vskip -2.5ex
      \caption{Illustration of the single port search mode in an 8T CAM.}
      \label{CAM}
      \vskip -2.5ex
  \end{figure}

\subsection{State-Matching CAM Optimization}
\begin{table}[t]
  \caption{Symbol Class and Alphabet Sizes with and without negation optimization (NO)}
\centering
\vskip -2ex
  \begin{adjustbox}{width=\columnwidth, center}
  \huge
  \begin{tabular}{cccccc}
  
    \hline
    
     \multirow{2}{*}{\textbf{Benchmark}} & {\textbf{Symbol}} & {\textbf{Symbol Class}} & {\textbf{Alphabet}} & {\textbf{\# CAM Entries with}} & {\textbf{\# CAM Entries}}\\
 {} & \textbf{Class Size} & \textbf{Size with NO} & \textbf{Size} & \textbf{Raw Symbol Class}&{\textbf{with NO}}\\
    \hline
    \hline
    Brill & 1&1&256&42658&42658 \\
    \hline
    ClamAV & 1.006&1.006&256&49593&49593\\
    \hline
    Dotstar & 1.56&1.56&256& 103280& 103280\\
    \hline
    Fermi & 7.18&4&256& 53769&61066\\
    \hline
    TCP & 9.26&1.28&256&32883&20156\\
    \hline
    Protomata & 4.41&2.65&256&162443&69715\\
    \hline
    Snort & 4.41&2.02&256&90718&72884\\
    \hline
    Hamming & 1&1&256& 11346& 11346\\
    \hline
    PowerEN & 1.95&1.09&256&48016&41080\\
    \hline
    Levenshtein & 1&1&256&2784&2784 \\
    \hline
    RandomForest &179.05&51.55&256&80515&75936\\
    \hline
    EntityResolution & 38.14&1.41&256&111996&95550\\
    \hline
    Bro217 & 1.55& 1.55&256&2352&2352\\
    \hline
    Dotstar03 & 1.92&1.3&256&14245&12445\\
    \hline
    Dotstar06 & 2.48&1.28&256&16536& 13116\\
    \hline
    Dotstar09 & 3.1&1.29&256&17834& 12723\\
    \hline
    Ranges1 & 1.29&1.29&115&12947 & 12947\\
    \hline
    Ranges05 & 1.21&1.21&107&12990&12990\\
    \hline
    SPM & 89.4&1.5&256&135675&100500\\
    \hline
    BlockRings & 1&1&2& 44352& 44352\\
    \hline
    ExactMath & 1.002&1.002&114& 12439& 12439\\
    \hline
  \end{tabular}
  
  \end{adjustbox}
  
  \label{table:symbol}
  \vskip -3ex
\end{table}

\begin{figure}[t]
      \centering
      \includegraphics[width=0.6\columnwidth]{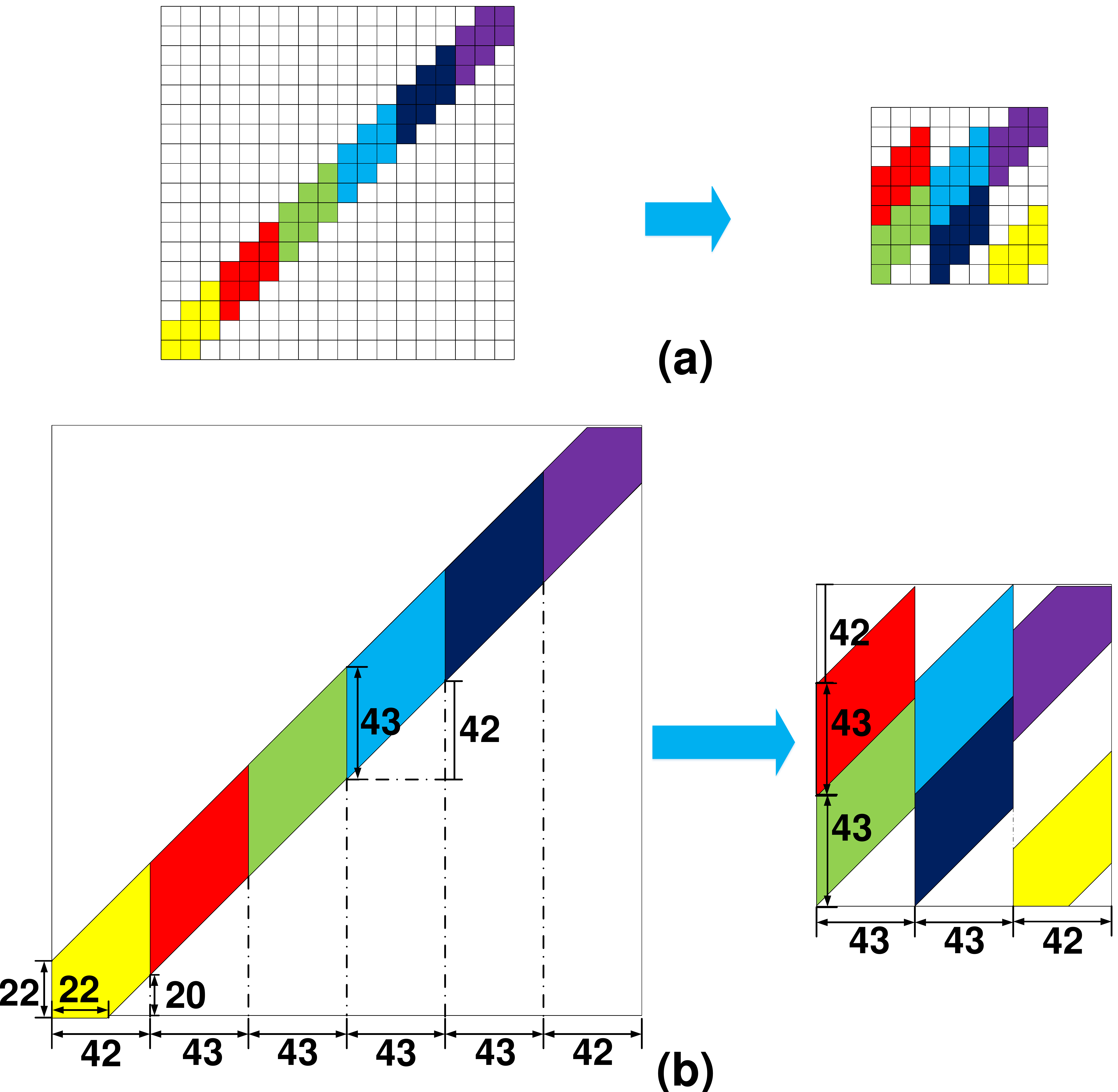}
      \vskip -2.5ex
      \caption{Sketch of the remapping (a) from 18$\times$18 FCB to 9$\times$9 RRCB and (b) from 256$\times$256 FCB to 128$\times$128 RRCB.}
      \label{rcbmap}
\vskip -3ex
  \end{figure}

\begin{figure}[t]
      \centering
      \includegraphics[width=0.88\columnwidth]{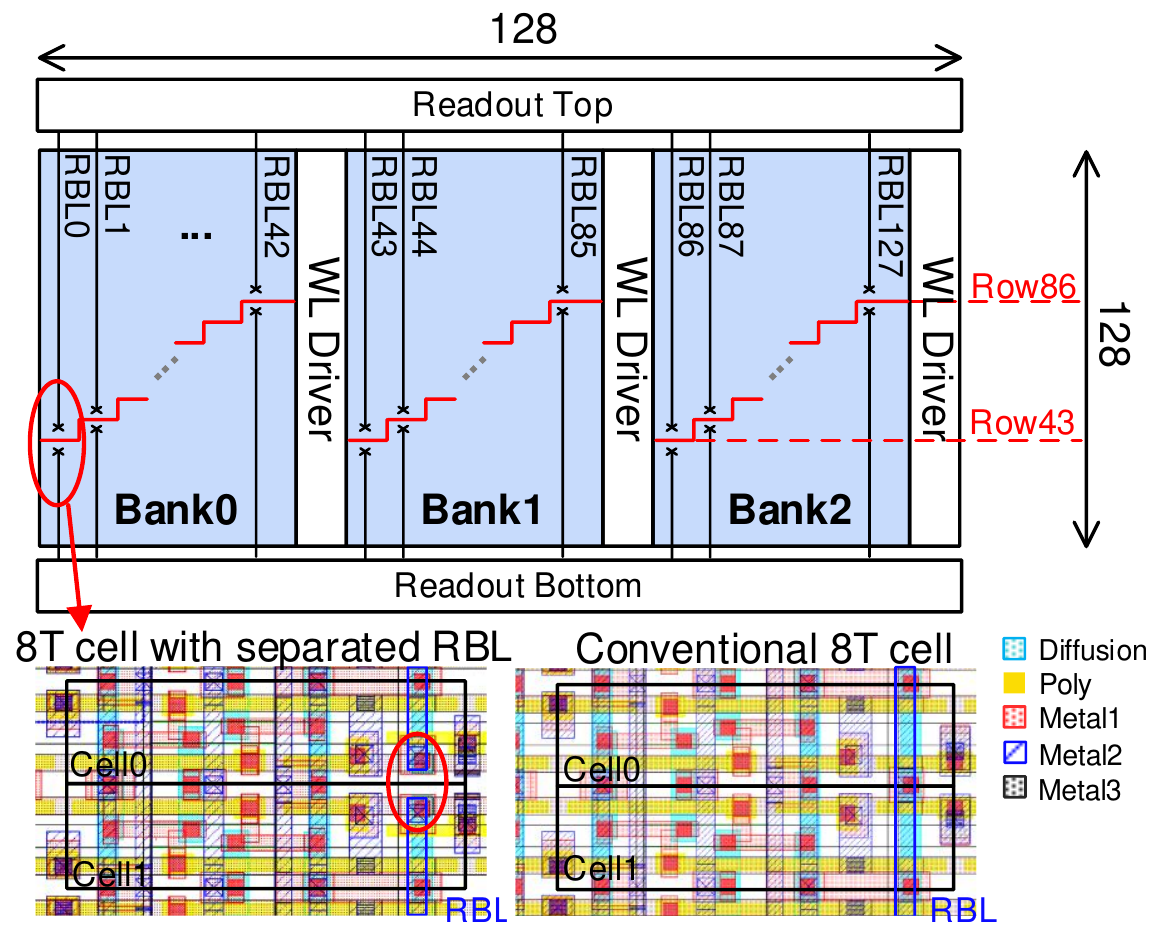}
      \vskip -2.5ex
      \caption{Implementation and layout of 128$\times$128 RCB.}
      \label{RCB}
\vskip -3ex
  \end{figure}
  
     \begin{figure*}[t]
      \centering
      \includegraphics[scale=0.55]{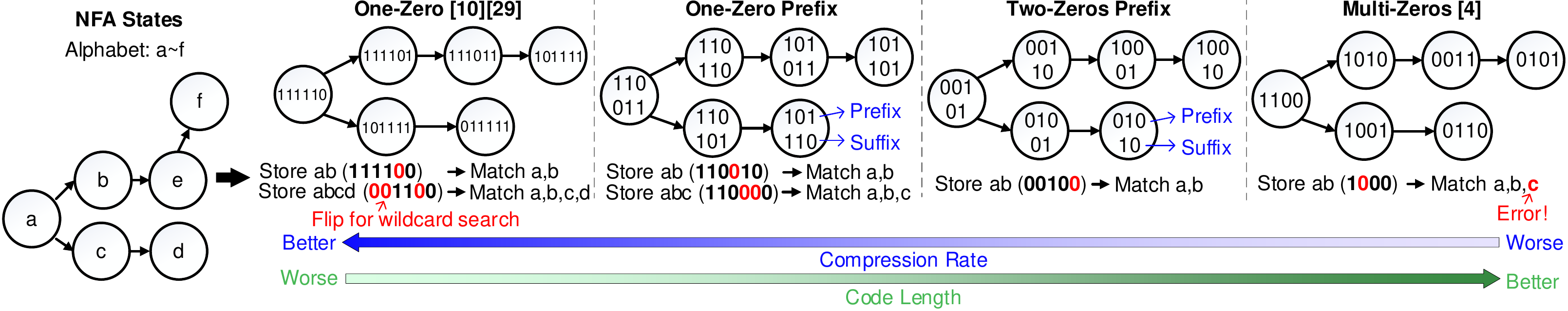}
      \vskip -2.5ex
      \caption{Illustrations of the encoding schemes.}
      \label{encoding}
      \vskip -3ex
  \end{figure*}
\par


In reducing the area and energy overhead of 16T (transistor) TCAM cells, \cite{jeloka_28_2016} showed that two standard 6T SRAMs can be configured for TCAM operations. We made a further innovation that a single search line (SL) is sufficient for the search operations if data and input symbols are encoded to have a fixed number of `1's. Therefore, standard push-rule 8T SRAM cells can be repurposed as 8T TCAM cells, but with only around 1/3 of the area~\cite{li_cam_2020}. As illustrated in Figure~\ref{CAM}, the read WL (RWL) and the read BL (RBL) in the 8T SRAM cell are repurposed as the SL and match line (ML) of CAM. Only when the cell stores a `1' and the search line is pulled up by input `1', a discharging path will be formed and report a mismatch. 
According to pigeonhole principle, when a fixed number of `1’ is enforced in both the stored and input symbols, any mismatch between the two symbols will lead to at least one discharging path and the result of mismatch. Any other types of symbols with varied number of `1's will result in reduced coding space and more complicated code compression~\cite{li_cam_2020}. 
However, TCAM mismatch should actually happen only when storing `1' and searching `0'. This can be achieved straightforwardly by adding inverters to all search line inputs, but a simpler method in CAMA is to directly embed the inversion into SRAM-based input encoder without any overhead, as required by the CAMA encoding scheme (see Section \S \uppercase\expandafter{\romannumeral5}). 
Since the search operation requires activating multiple SLs simultaneously, the proposed 8T cell with decoupled read port ensures reliable non-destructive research while the 6T cell based CAM~\cite{jeloka_28_2016} faces severe read disturbance issue, especially in large arrays and advanced technology nodes. It is worth mentioning that other coding schemes like ASCII do not work on the 8T CAM because of the special encoding match condition, e.g. the ASCII `A'(01000001) will wrongly match the ASCII `C'(01000011).
\par
Besides, some ANML-NFAs have symbol classes defined by negation $\widehat{ }~$, which can be implemented by matching the excluded symbols and inverting the output if this reduces the total number of CAM entries and CAM area. An inverter is added to each row of the state-matching CAM, which will be enabled when a symbol class defined by negation is met. 
For example, if one state’s symbol class is [~$\widehat{}~$abcd], it will accept all the other 252 symbols except a, b, c and d and its symbol class size is 252 but the size of its negation symbol class [abcd] is 4.
This is named Negation Optimization (NO). As Table~\ref{table:symbol} shows, NO reduces the number of CAM states in nearly all applications, especially in cases where the original symbol class size is far larger than the optimized symbol class size after NO. 

\par
The column precharger at each CAM entry is implemented to selectively enable entries to save energy consumption. It functions as the AND gates between the transition results and match results during state transition. Meanwhile, the CAM bank contains a mask so that it can turn off certain search bits to save energy when the bit width of search data is smaller than the CAM's word length, i.e. the number of rows.

\par

\subsection{Switch Network Optimization}

The proposed 128$\times$128 RRCB remaps the diagonal transitions in the 256$\times$256 FCB to achieve a dense data storage. Figure~\ref{rcbmap}(a) illustrates a toy example of the remapping where the diagonal transitions are divided into groups in FCB. The neighboring groups are remapped to stacking into the same columns in RRCB to save area. To fit into the $9\times 9$ RRCB, the diagonal width $k_{dia}$ that each group occupies is chosen as 3. In reality, see Figure~\ref{rcbmap}(b) we choose $k_{dia}=43$ to fit into the 128$\times$128 RRCB and ensure only two groups are stacked in each column. Meanwhile, the WLs are divided into 3 segments with each one receiving different inputs. As shown in Figure~\ref{RCB}, it requires a split RBL in each column and two read peripheries on the top and bottom. Compared to the layout of conventional 8T cell (see Figure~\ref{RCB}), the proposed 8T cell with split RBL simply cuts off the shared drain of NMOS and will not introduce area overhead. As a comparison, eAP chooses $k_{dia}=21$ for the 96$\times$96 RCB, leading to 3 segments of BLs in each column which is impractical for the physical implementation of the readout peripheries.

\par

 For dense transitions that RCB cannot handle, RRCB can be reconfigured into a 128$\times$128 FCB by combining the input and output data. The two split RBLs can be considered as one once the readout data from the top and bottom are combined together using OR gates. Similarly, feeding three replicas of inputs to each WL segment is equivalent to connecting the WLs in the same row. As a result, there is no need to reuse the state-matching memory like eAP, relaxing the storage requirement of  CAM.

\par

There exist some situations that the CCs are too large to fit into one CAM bank. Thus, global switches are used to provide connections between different CAM banks. We adopt the 256$\times$256 8T SRAM bank similar to eAP \cite{sadredini_eap_2019}. Each local switch allows 16 STEs sending to the global switch and 16 STEs receiving from the global switch.

%% file: S5.tex
\section{Data Encoding Schemes and Optimization}

\subsection{Encoding Scheme}
\par
\par
\par
\par
\par
The goal of optimizing the encoding scheme is to achieve an optimal trade-off between the code length and the compression rate. Since NFA state's symbol class size will sometimes be larger than 1, we need to compress multiple symbols into one code while maintaining moderate coding complexity. The matching condition of the proposed 8T CAM, as discussed in Section \S \uppercase\expandafter{\romannumeral4}.A, is as follows: while `1' in the cell only matches `1' inputs, `0' represents a \textit{don't care} that matches both `0' and `1' inputs. Therefore, compression is achieved by flipping certain `1's into `0's in the codes. E.g. states `a' and `b' are encoded as `0111' and `1011', respectively, so `ab' is encoded as `0011' to match both symbols. Meanwhile, the number of zeros should be fixed for all symbols with class size equal to 1, as mentioned in \S \uppercase\expandafter{\romannumeral4}.A. As a result, the unique matching condition and the encoding constraints require specialized encoding strategies.

\par
We observe that the most important design choice of the encoding is the number of zeros in the codes because it determines both the code length and the compression rate. The \textit{One-Zero} encoding \cite{subramaniyan_cache_2017,dlugosch_efficient_2014}, which has fixed one `0' in the codes, achieves the largest compression rate because any combinations of the symbols can be compressed into one code by flipping `1's at any positions. As Figure~\ref{encoding} shows, states `a'/`b'/`c'/`d' can be simply compressed into one code. However, its code length, which is equal to the alphabet size, is overly large and thus causes low memory utilization. 

On the contrary, an equal number of `1's and `0's leads to the shortest code length but the worst compression rate. This is termed as \textit{Multi-Zeros} encoding~\cite{li_cam_2020}. As shown in Figure~\ref{encoding}, a wildcard search for `ab' by flipping the second `1' will cause a wrong matching of state `c' due to the poor compression ability. Quantitatively, the numbers of symbols ($N_S$) that can be compressed into one code must follow the \textit{combinatorial number rule}: $N_S=\tbinom{m}{n}$, where $m$ is the possible number of `0's in the code after compression while $n$ is the number of `0's in the original code. In this example, $m=3$ or $4$ and $n=2$, and thus $N_S=3$ or $6$. Therefore, the $2$ ($\neq N_S$) states `a' and `b' cannot be compressed into one code. Generally, the more ‘0’s one code has, the more discrete the combinatorial numbers will be, leading to increased difficulty of compression and high state numbers. 
\par
To address the trade-off, we combine these two encoding schemes above and divide the code into two parts: the prefix and the suffix (see Figure~\ref{encoding}). The suffix follows the One-Zero encoding style to achieve the maximum compression capability. On the other hand, the overall code length is shortened by adjusting the number of `0's in the prefix at the cost of less compression space.

\par
\par
The first scheme, namely \textit{Two-Zeros Prefix} encoding, has fixed two `0's in its prefix for each symbol (see Figure~\ref{encoding}). By flipping the `1's in the suffix, symbols with the same prefix can be compressed into one code. This is termed as \textit{suffix compression}. For example, if state `a' is `001 01' and `b' is `001 10', state `ab' will be encoded to `001 00' (see Figure~\ref{encoding}). Similarly, a \textit{prefix compression} is also proposed to flip `1's in the prefix and compress the codes with the same suffix (not shown in the figure). As mentioned above, the prefix compression has more restricted compression space since two `0's exist in the prefix but achieves smaller code length.
\par
The second scheme is the \textit{One-Zero Prefix} encoding where the prefix has fixed one `0'. Compared to the previous scheme, its prefix compression has better compression rate: any number of codes within the same suffix can be compressed to one code because there is only one `0' in the prefix. For instance, if states `a'/`b'/`c' are encoded as `110 011'/`110 110'/`110 101', state `abc' can be compressed into `110 000' (see Figure~\ref{encoding}). Despite the improved compression capability, the code length becomes larger and thus it is specialized for those applications with large average symbol class sizes. 
\par
\subsection{Encoding Selection and Symbol Clustering}
We observe that using a single encoding scheme for all benchmarks is inefficient. E.g., using a 32-bit One-Zero prefix encoding scheme to encode the BlockRings\cite{wadden_anmlzoo_2016} is a waste, as 32-bit codes can express 256 symbols but BlockRings only has 2 symbols. 
Thus, an encoding selection algorithm is proposed to automatically select a proper encoding scheme by aiming to find an encoding with minimum code length and adequate compression space to balance the number of CAM states and the code length.


\par
First, we calculate the alphabet size ($\mathscr{A}$) of ANML/MNRL dataset and the average symbol class size with NO ($\mathscr{S}$). 
If $\mathscr{S}$ is equal to 1, the Multi-Zeros code is directly selected as it requires no compression. Its code length is calculated by 
\begin{gather}
    \mathscr{L}=minimize\{\mathscr{L}:\tbinom{\mathscr{L}}{\frac{\mathscr{L}}{2}}\geq\mathscr{A}\}
\end{gather}
Otherwise we need choose from Two-Zeros prefix and One-Zero prefix encoding schemes based on the code length. For Two-Zeros prefix encoding, the length of the suffix is set to be larger than the symbol class size so that more symbols can be compressed into one CAM entry. By sweeping the length of the suffix from $\mathscr{S}$ to $\sqrt{\mathscr{A}}$, the minimum code length $\mathscr{L}_{min}$ can be calculated based on $\mathscr{A}$: 
\begin{gather}
    \mathscr{L}_{min}=min(\mathscr{L}_{min}, l_{s}+minimize\{l_{p}:\tbinom{l_{p}}{2} \times l_{s}\geq\mathscr{A}\})
\end{gather}
where $l_{s}$ and $l_{p}$ are the length of suffix and prefix, respectively. For example, given $\mathscr{S}=5$ and $\mathscr{A}=256$, the length of Two-Zeros prefix encoding is 16b because when the suffix length $l_{s}$ is set from 5 to 16, the minimal $\mathscr{L}$ that satisfies $C_{l_{p}}^{2}\times l_{s}\geq256$ is 16. For One-Zero prefix encoding, the shortest code length is 2$\sqrt{\mathscr{A}}$ according to Cauchy's inequality. Therefore, by comparing 2$\sqrt{\mathscr{A}}$ and $L_{min}$, the encoding with smaller code length will be selected. Specially, if the alphabet is small enough, which means all states can be directly stored into one CAM entry, the One-Zero encoding will be adopted. 

\par
After the encoding scheme is determined, the symbol clustering algorithm groups the symbols that have a high probability to occur in the same symbol class into one cluster with the same prefix (frequency-first), leading to the reduced number of CAM states. Meanwhile, this algorithm ensures most of the symbols can be compressed with suffix compression, which simplifies the compression procedure because of the decreased demands for prefix compression. At the beginning of the algorithm, we calculate the frequency for each symbol. For each unfilled cluster, we estimate the possibility that the undistributed symbols and the symbols in the cluster appear in the same symbol class, the P($\mathscr{X}\mathscr{C}$) ($\mathscr{C}$: symbols in chosen cluster, $\mathscr{X}$: the undistributed symbol). The symbol with the highest probability will be packed into that cluster. The clustering operation repeats until all clusters are full.

\par
\subsection{Encoding Results Analysis}
Table~\ref{table:coding} shows the proposed encoding scheme with optimization only increases the average number of states by 13$\%$ compared to one-hot encoding. Without code selection and clustering optimization, the 32-bit One-Zero prefix encoding not only needs longer code length but also increases the states by 25$\%$. The relatively large state increasing (about 1.23$\times$) in RandomForest is because its symbol class size with NO is more than 50 (Table~\ref{table:symbol}). Overall, the proposed encoding scheme and optimization algorithm successfully map all benchmarks into CAMA without significance state increases, leading to the area and energy efficiency boosts.
\begin{table}[t]
  \caption{Comparison of CAMA encoding scheme with 256-bit and fixed 32-bit One-Zero prefix Encoding without clustering optimization (Memory Usage = Code Length $\times$ \# States)}
  \centering
  \vskip -2ex
  \begin{adjustbox}{width=0.44\textwidth, center}
  \begin{tabular}{ccccc}
    \hline
    {}&{}&{}&\multicolumn{2}{c}{{\textbf{Proposed}}}\\
    \cline{4-5}
    {\textbf{Benchmark}}&{\textbf{256-bit One-}}&{\textbf{Fixed 32-bit One-Zero}}&{\textbf{Code}} & {\textbf{Encoding}}    \\
  {}&{\textbf{Zero States}}&{\textbf{Prefix Encoding States}}&{\textbf{Length}} & {\textbf{States}} \\
    \hline
    \hline
    Brill &42658&42658& 11&42658 \\
    \hline
    ClamAV &49538&49616& 16&49593\\
    \hline
    Dotstar &96438&99254& 16&103280\\
    \hline
    Fermi &40783&43972& 16&61066\\
    \hline
    TCP &19704&20200& 16&20156\\
    \hline
    Protomata &42011&78078& 16&69715\\
    \hline
    Snort &69029&88857& 16&72884\\
    \hline
    Hamming&11346&11346 & 11&11346\\
    \hline
    PowerEN &40513&41511& 16&41080\\
    \hline
    Levenshtein &2784&2784& 11&2784 \\
    \hline
    RandomForest &33220&128451&32&75936\\
    \hline
    EntityResolution &95136&139994& 16&95550\\
    \hline
    Bro217 &2312&2312& 16& 2352\\
    \hline
    Dotstar03 &12144&12325& 16&12445\\
    \hline
    Dotstar06 &12640&12874& 16&13116\\
    \hline
    Dotstar09 &12431&13000& 16&12723\\
    \hline
    Ranges1 &12464&12645& 13&12947\\
    \hline
    Ranges05 &12439&12801& 12&12990\\
    \hline
    SPM &100500&130650& 16&100500\\
    \hline
    BlockRings &44352&44352& 2&44352\\
    \hline
    ExactMath &12439&12451& 16&12439\\
    \hline
  \end{tabular}
  \end{adjustbox}
  \label{table:coding}
 \vskip -4ex
\end{table}

{The worst case of state increasing will only occur in One-Zero prefix encoding with 2$\sqrt{\mathscr{A}}$ code length because the worst case is supposed to have the maximum number of CAM entries with large symbol class size. Theoretically, the upper bound of One-Zero prefix for one state is $2\mathscr{A}$ memory bits ($=$number of entries$\times$number of bits in one state) in state matching which only occurs when any two symbols in a symbol class have different suffixes and prefixes. However, having a few such symbol classes will trigger re-clustering of symbols, unless in the rare case where the distribution of symbols and symbol classes is truly random. Overall, it is unlikely to reach the worst case that requires $2\mathscr{A}$ memory bits. For example, the maximum average number of state matching memory bits for one state is only 73 among our benchmarks, which happens in RandomForest with dense transitions. }

%
\par

%% file: S6.tex
\section{System Implementation}
\subsection{Overall Design}

\begin{figure}[t]
      \centering
      \includegraphics[scale=0.2]{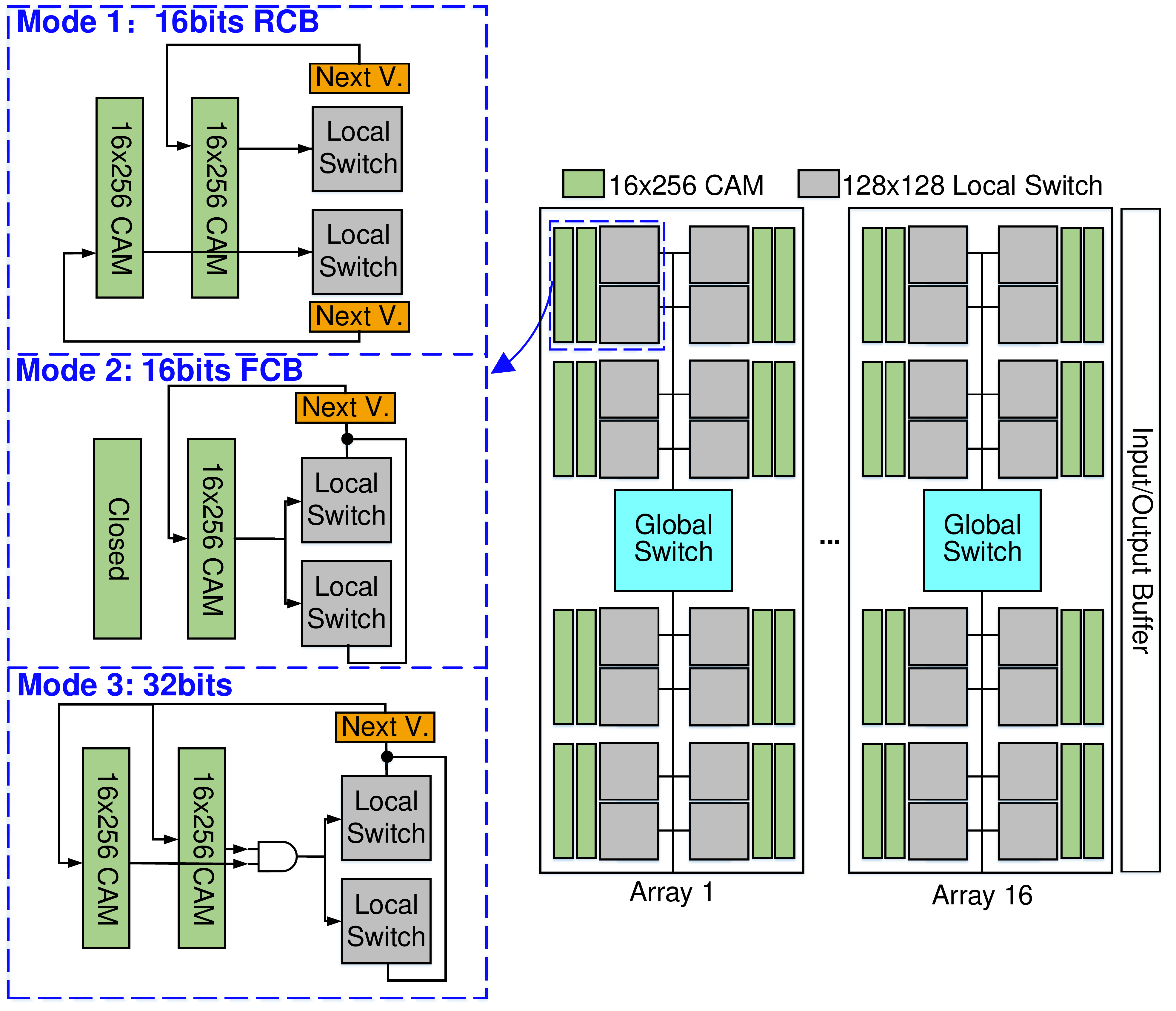}
      \vskip -2.5ex
      \caption{General overview of bank abstraction and three different modes.}
      \label{mode}
  \vskip -4ex
  \end{figure}

The system diagram of a CAMA bank is shown in Figure~\ref{mode}, which can store up to 65536 states. Each bank consists of 16 arrays with 8 tiles in each array sharing one global switch.
Inside the tile, two 128$\times$128 local switches are stacked vertically to match the length of the two horizontally placed 16$\times$256 CAM sub-arrays, leading to a compact physical layout. The number of CAM entries is chosen based on the encoding results in Table~\ref{table:coding} where 16 is the maximum value in all benchmarks except RandomForest. 
All the encoded inputs are stored in a 256$\times$32 SRAM (not drawn in the figure) which only consumes only 0.1\% of total energy on average.

According to the circuit and algorithm design, there exists some cases where we have to use 32-bit one-0 prefix encoding for larger compression space or store the dense transitions in local switches that are reconfigured to FCB. Therefore, three processing modes are designed to handle different situations: 16-bit RCB mode, 16-bit FCB mode, and 32-bit mode (see Figure~\ref{mode}). Those modes cover nearly all applications with the alphabet size no larger than 256 (the largest alphabet size is 256 in ANMLZoo and Regex) since the longest code length of CAMA is 32 bit using One-Zero prefix encoding. 

\begin{figure*}[tb]
      \centering
      \includegraphics[scale=0.22]{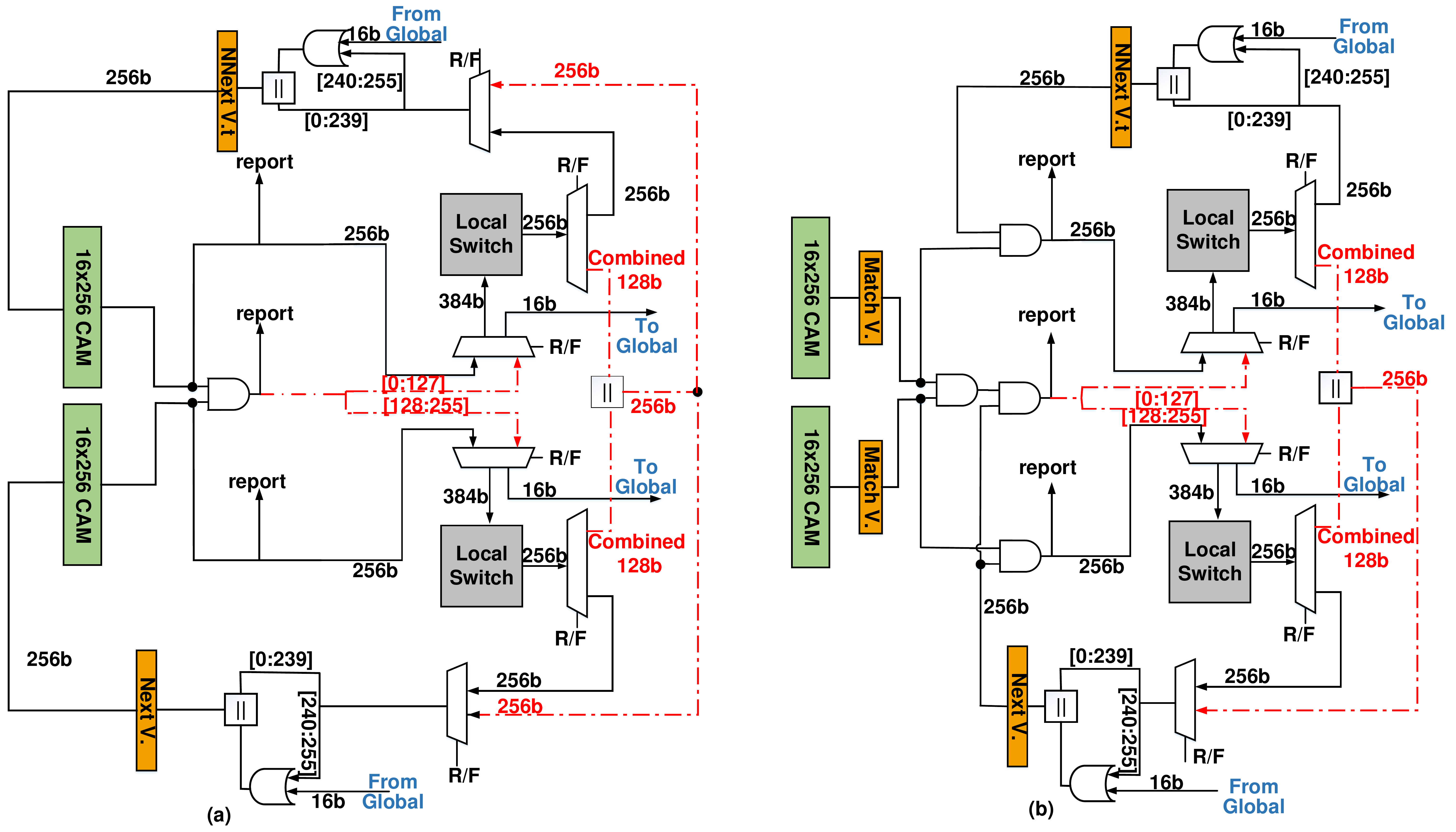}
      \vskip -1.5ex
      \caption{Detailed implementation of one tile in CAMA: (a)CAMA-E with selective row enabling and (b) CAMA-T with pipelined matching and transition for higher throughput. }
      \label{circuit}
      \vskip -4ex
  \end{figure*}
The 16-bit RCB mode is activated when the application can be expressed by codes less than 16-bit and the interconnection can be stored into the 128$\times$128 RCB-mode local switch. In this mode, both CAM sub-arrays in the tile are connected to the local switches
(see Mode 1 in Figure~\ref{mode}). Figure~\ref{circuit} illustrates the detailed implementation of each tile, where the connection of 16-bit mode is drawn with solid lines. As mentioned in \S \uppercase\expandafter{\romannumeral4}.B, the 128$\times$128 RRCB has 3 WLs in each row and 2 BLs in each column, resulting in the 384-bit input and 256-bit output. After the state matching, the MUX before RRCB's input will choose the 256-bit CAM output and replicate some of them to 
match the 384-bit input bus (The replica connection is fixed). The deMUX after RRCB's readout routes 256-bit output to the Next Vector register. This mode has the lowest energy consumption and the highest area utilization. 
\par
In the 16-bit FCB mode, the system utilizes an encoding scheme whose code length is less than 16-bit but the local switches in RCB mode is not able to store all the transitions. Each tile turns off one of the CAM sub-arrays by power-gating and connects the two local switches to the other sub-array as shown in the Mode 2 in Figure~\ref{mode}. 
The working CAM is divided into two 16$\times$128 parts conceptually, and each local switch is operating as a normal 128$\times$128 FCB SRAM bank to provide a full 128$\times$128 interconnection for the two parts. As shown by the dash lines in Figure~\ref{circuit}, 256-bit ANDed results from the two CAM sub-arrays (one of sub-arrays is turned off but its results are reset to `1's) are separated into two 128-bit parts and connected with 384-bit input through the MUX after replication (the replica is fixed) while the readout results of RRCB are combined (OR) to 128-bit data after the deMUX. Finally, the combined outputs of two RRCBs are merged to the 256-bit output bus and sent back to the Next Vector register.
\par
The system is configured to 32-bit mode when the 16-bit encoding cannot cover large symbol classes of applications.
As described by Mode 3 in Figure~\ref{mode}, match results from the two sub-arrays are combined by an AND gate and then sent to the two local switches. Its circuit level working principle is just like the 16-bit FCB mode except the code length and the number of working CAMs.
Although this mode has the max compression for symbol classes, this 32-bit mode has the worst area utilization and consumes the highest power.

\subsection{Pipeline, Input and Output}
\begin{figure}[t]
      \centering
      \includegraphics[scale=0.3]{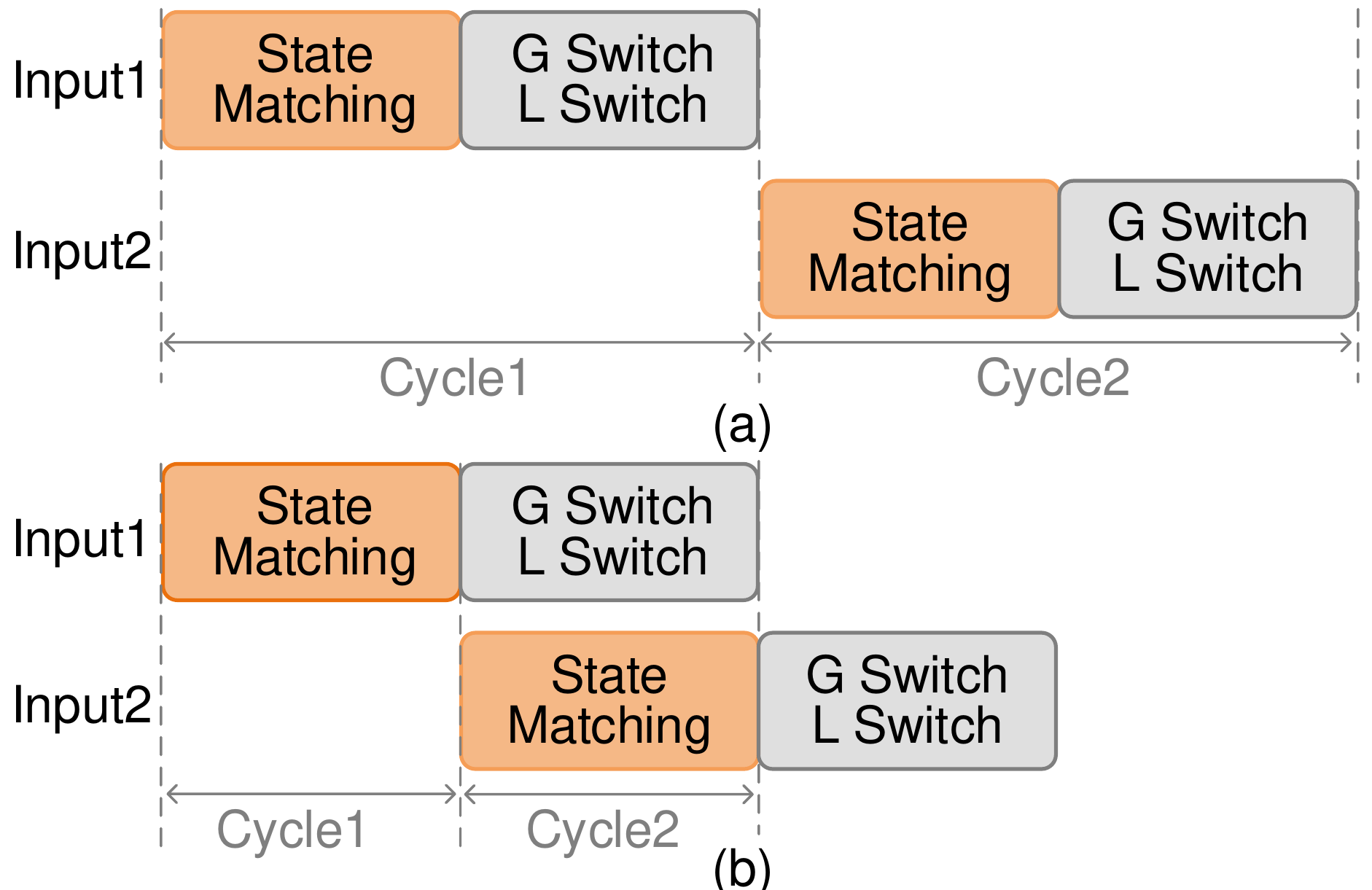}
      \vskip -2.5ex
      \caption{Timing diagram of (a) non-pipeline and (b) pipeline implementations.}
      \label{pipeline}
        \vskip -3ex
  \end{figure}

There are two versions of CAMA, namely CAMA-E and CAMA-T, targeting high energy efficiency and high throughput, respectively. CAMA-E saves energy by only precharging the enabled STEs during state transition (Figure~\ref{circuit}(a)), where the transition results are directly sent back to CAM and the column precharger performs the AND operation. However, CAMA-E cannot pipeline state matching and state transition because the next symbol must obtain the transition results from the last symbol but the direct feedback causes conflicts. As shown in Figure~\ref{circuit}(a), the non-pipelined architecture leads to relatively low clock frequency. On the other hand, CAMA-T, a configuration optimized for throughput, connects transition results to AND gates instead of feeding back them to the CAM to realize the pipeline as shown in the Figure~\ref{circuit}(b) and Figure~\ref{pipeline}(b). In the state matching phase, the results are stored in the Match Vector register which acts as the pipeline register. The rest of the operations are similar to CA. Since the trade-off between the energy and throughput is application-dependent, we evaluate both non-pipelined (CAMA-E) and pipelined (CAMA-T) in the experiments.

As for the data storage, CAMA uses two buffers to store the input symbols and output report vectors. The input symbols are stored in the 128-entry input buffer similar to CA. Every cycle one symbol is deleted and distributed to all arrays through the bus. Every time the input buffer is empty, the buffer will cause an interruption and get a set of new symbols. 
\par
A 64-entry output buffer is designed to keep a similar interruption frequency between input and output. Similar to input interruption, every time the output buffer is full, an interruption is sent to the CPU, and CPU will get the reports and clear all entries. \cite{wadden_characterizing_2018} characterizes the reporting behavior on ANMLZoo, showing that 10 out of 12 benchmarks have less than 0.5 reports per cycle on average on 1MB input. 
Therefore, this ratio inspires us to adopt the 64-entry output buffer to hide the output interruption behind input interruption. Each entry stores the matching results and the corresponding reporting state ID. The mapping of the report states is realized by setting up a report mask for each bank. Once a report is detected, the bank will create an output entry and store the active state ID, partition ID, input symbol, and the reporting cycle.

%% file: S7.tex
\section{Evaluation Methodology}
\par
\noindent\textbf{NFA workloads: }We evaluate the proposed architecture and algorithmic optimization by using ANMLZoo\cite{wadden_anmlzoo_2016} and Regex\cite{becchi_workload_2008} benchmarks. These benchmarks cover a wide range of real-world applications, including network intrusion detection\cite{becchi_workload_2008}, machine learning, gene sequence matching\cite{roy_discovering_2016} and data mining\cite{wang_pattern_mining_2016}. While a more recent benchmark suite AutomataZoo\cite{wadden_automatazoo_2018} extends ANMLZoo with larger benchmarks. ANMLZoo is selected in our evaluation because (1) the AP chip with up to 48k states only supports ANMLZoo, (2) evaluating an automata accelerator on large benchmarks in AutomataZoo is excessively time consuming, and (3) 
prior studies on automata acceleration, including CA~\cite{subramaniyan_cache_2017}, Impala~\cite{sadredini_impala_2020}, AP~\cite{dlugosch_efficient_2014}, FlexAmata~\cite{sadredini_flexamata_2020}, and Grapefruit~\cite{rahimi_grapefruit_2020}, all adopt ANMLZoo for evaluation. 10MB inputs are used for all evaluations. 
\par
\noindent\textbf{Experiment Setup: }The open-source virtual automata simulator VASim~\cite{wadden_anmlzoo_2016} is used to evaluate CAMA and prior designs. VASim is modified to support the proposed encoding scheme. The simulator takes in one ANML or MNRL file, and calculates matches and transitions of each input symbol cycle by cycle and provides cycle-accurate energy estimation of the archtectures.

Table~\ref{table:symbol-set} lists memory parameters in our evaluations, including access energy, delay and area. These numbers come from SPICE simulations on custom designed SRAM and CAM arrays in TSMC 28nm CMOS. The 6T SRAM parameters at different sizes and configurations are very close to 6T SRAMs from TSMC compilers. We cannot entirely rely on the compiler because it does not offer some of the memory types and configurations in our evaluations. For fair comparison, we re-simulate SOTA architectures with the same circuit models. 

\par

\begin{table}[t]
\vskip -1.5ex
  \caption{Circuit models in 28nm}
\centering
\vskip -2.5ex
  \begin{adjustbox}{width=\columnwidth, center}
  \huge
  \begin{tabular}{c|ccccc}
  
    \hline
    
     {\textbf{Type}} & {\textbf{Size}} & {\textbf{Energy (pJ)}} & {\textbf{Delay (ps)}} & {\textbf{Area ($\mu$m$^2$)}}& {\textbf{Leakage ($\mu$A)}} \\
    \hline
    \hline
    
    \multirow{2}{*}{6T SRAM} 
    
     & 256$\times$256&19.45&416&14877&532\\
    \cline{2-6}
     & 16$\times$256& 15.3&317&3659&247\\
    \hline
    
     \multirow{2}{*}{8T SRAM} & 128$\times$128 & 8.67& 292& 5655&243 \\
    \cline{2-6}
    
    & 256$\times$256&17.9&394&18153&584\\
    \hline
    8T CAM & 16$\times$256&16.78& 325& 3919&299\\
    \hline
  
  \end{tabular}
  
  \end{adjustbox}
  
  \label{table:symbol-set}
  \vskip -3ex
\end{table}

\section{Results}
In this section, we first compare the throughput of existing in-memory automata processors, followed by analysis of area efficiency, compute density and energy/power consumption. 

\subsection{Overall Performance}

The overall performance of the spatial automata processing architectures is defined as the clock period or throughput (frequency$\times$bits/cycle). As Table~\ref{table:delay} shows, CAMA and 2-stride Impala’s state matching has less delay than CA and eAP. This is because both designs have a denser state matching array and reduce the length of BLs from 256 to 16. Shorter BLs result in less RC delay. Benefiting from the compact local switch design, CAMA and eAP will have less delay in local switch. Because of the long-distance routing in the global switch, the delay consists of two parts: the wire delay and the memory access delay. Based on CA's model, the wire delay on global routing is 99ps. As a result, the global delay for CA is 493ps with 394ps read delay and 99ps wire delay in 28nm. Since 2-stride Impala has 2$\times$ less state matching area overhead than CA and CAMA is 4$\times$ less than CA (area listed in Table~\ref{table:symbol-set}), the wire delay of 2-stride Impala and CAMA is estimated as 48.69ps and 26.1ps, repectively. Due to the large area of 8T SRAM in state matching, eAP has the largest wire delay of 121ps. 
It is worth mentioning that the original three-stage pipeline used in CA will cause a data-hazard issue \cite{sadredini_eap_2019}. Thus, we adjust the original CA pipeline and make the local switch and global switch process simultaneously, similar to the settings in \cite{sadredini_eap_2019}. 
\par 
\begin{table}[t]
  \caption{Delays and frequency in 28nm}
\centering
\vskip -2.5ex
  \begin{adjustbox}{width=0.48\textwidth, center}
  \begin{tabular}{c|ccccc}
  
    \hline

     {\textbf{Design}} & {\textbf{State Match}} & {\textbf{L-switch}} & {\textbf{G-switch}} & {\textbf{Freq.Max}}&{\textbf{Freq.Operated}} \\
    \hline
    \hline
    
    CAMA-E&325ps&	292ps&	420.1ps&1.34GHz	&1.21GHz\\
    \hline
    CAMA-T&325ps&	292ps&	420.1ps&2.38GHz&	2.14GHz\\
     \hline
    2-stride Impala~\cite{sadredini_impala_2020}&317ps&394ps&442.69ps&2.26GHz&2.03GHz\\
    \hline
eAP~\cite{sadredini_eap_2019}&394ps&394ps&515ps&1.94GHz&1.75GHz\\
\hline
CA~\cite{subramaniyan_cache_2017}&416ps&394ps&493ps&2.03GHz&1.82GHz\\
\hline
AP~\cite{dlugosch_efficient_2014}&N/A&N/A&N/A&0.133GHz&0.133GHz\\
   
    \hline
  
  \end{tabular}
  
  \end{adjustbox}
  
  \label{table:delay}
  \vskip -3ex
\end{table}
\par
The largest stage latency determines the frequency of the pipeline. Designs of 2-stride Impala, CA and CAMA-T all have the slowest delay on global switches. The frequency of CAMA-E (the non-pipeline design) is bottlenecked by the state matching delay plus the global switch delay while the local switch delay is hidden behind the global switch delay because they are working in parallel. We choose a 10$\%$ lower frequency than the max frequency to provide margin for estimation errors. 
Overall, CAMA-T has higher throughput than CA, 2-stride Impala, eAP and CAMA-E by 1.18$\times$, 1.05$\times$, 1.22$\times$ and 1.76$\times$ in 28nm, respectively. This is mainly because CAMA has smaller state matching array leading to smaller global wire delay which is the bottleneck for all designs. CAMA-T and CAMA-E provide a 16.1$\times$ and 9.1$\times$ speedup than AP(50nm).

\begin{table}[t]
\caption{Switch mapping results for CA and CAMA}
  \centering
 
  \vskip -2.5ex
  \begin{adjustbox}{width=0.47\textwidth, center}
  \begin{tabular}{cccccc}
    \hline
    
    \multirow{2}{*}{\textbf{Benchmark}} & {\textbf{Baseline}} & {\textbf{Baseline}} & {\textbf{Proposed}} & {\textbf{Proposed}} &  {\textbf{Proposed}} \\
 {} & \textbf{Local} & \textbf{Global} & \textbf{RCB mode} & \textbf{Global} & \textbf{FCB mode}\\
    \hline
    \hline
    Brill & 169&0&169 &0&0 \\
    \hline
    ClamAV & 199&3&199&0&3\\
    \hline
    Dotstar & 381&0&408&0&0\\
    \hline
    Fermi & 160&0&245&0&0\\
    \hline
    TCP & 78&1&76&1&8\\
    \hline
    Protomata & 166&0&274&1&5\\
    \hline
    Snort & 277&0&284&1&27\\
    \hline
    Hamming & 47&0&47&0&0\\
    \hline
    PowerEN & 162&0&162&0&0\\
    \hline
    Levenshtein & 12&0&12&0&0 \\
    \hline
    RandomForest &139&0&0&40&662\\
    \hline
    EntityResolution & 500&0&0&0&1000\\
    \hline
    Bro217 & 10& 0&10&0&0\\
    \hline
    Dotstar03 & 49&0&50&0&0\\
    \hline
    Dotstar06 & 51&0&53&0&0\\
    \hline
    Dotstar09 & 50&0&51&0&0\\
    \hline
    Ranges1 & 50&0&52&0&0\\
    \hline
    Ranges05 & 51&0&53&0&0\\
    \hline
    SPM & 419&0&419&0&0\\
    \hline
    BlockRings & 192&0&192&0&0\\
    \hline
    ExactMath & 50&0&50&0&0\\
    \hline
  \end{tabular}
  \end{adjustbox}
  \label{table:mapping}
  \vskip -3ex
\end{table}
\subsection{Area Overhead and Compute Density}

CAMA-T and CAMA-E benefit from the area reduction in both state matching phase and state transition phase while Impala only optimizes the state matching and eAP only reduces the memory area of state transition. CA has the largest area overhead since neither of the phases are optimized. Table~\ref{table:mapping} shows the mapping result in CAMA, where CAMA-T and CAMA-E have the same mapping. Overall, CAMA has lower area overhead than 2-stride Impala, eAP and CA by 1.91$\times$, 1.78$\times$ and 2.48$\times$, respectively. 
\par
\begin{figure}[t]
      \centering
      \includegraphics[width=0.92\columnwidth]{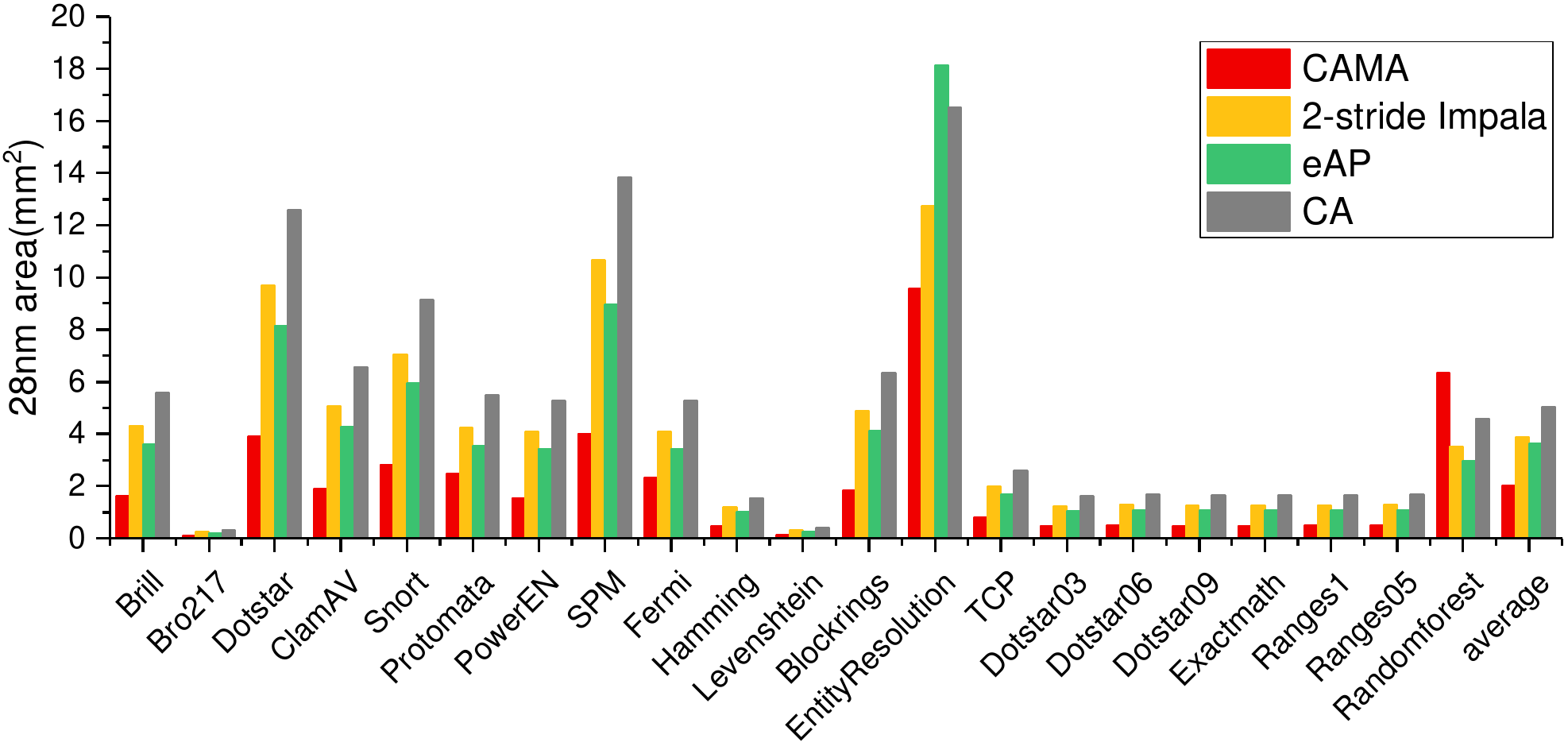}
      \vskip -2.5ex
      \caption{Area comparison.}
      \label{area}
      \vskip -3ex
  \end{figure}
Due to the faster clock frequency and area optimization in both phases, CAMA features the highest compute density which is defined as throughput per unit area (TOPS/mm$^2$). CAMA-T provides an average 2.68$\times$, 3.87$\times$ and 2.62$\times$ higher compute density than 2-stride Impala, CA and eAP in 28nm. Meanwhile, even if CAMA-E has smaller throughput, it shows an average 1.51$\times$, 2.19$\times$ and 1.48$\times$ higher compute density than 2-stride Impala and CA, eAP thanks to the reduced memory area. Only a few applications, e.g., EntityResoluton and RandomForest which cannot fit into RCB-mode switch, show lower compute density as it works in the 16-bit FCB mode where only one CAM sub-array in the tile is operating.

\subsection{Energy/Power Consumption}

To calculate the energy consumption in each cycle, we need to know the number of active partitions, the number of enabled partitions which is the key factor in CAMA-E (can be overlooked by other pipelined designs), the number of active rows in both state matching and state transition phases, and the number of dynamic transitions between partitions which result in increased global switch accesses and wire energy. All these factors are influenced by the mapping algorithm. We use a greedy mapping algorithm similar to CA to pack the connected components into hardware, and thus the number of active partitions are reduced.

\begin{figure}[!htb]
  \centering
  \vskip -1ex
      \includegraphics[scale=1.44]{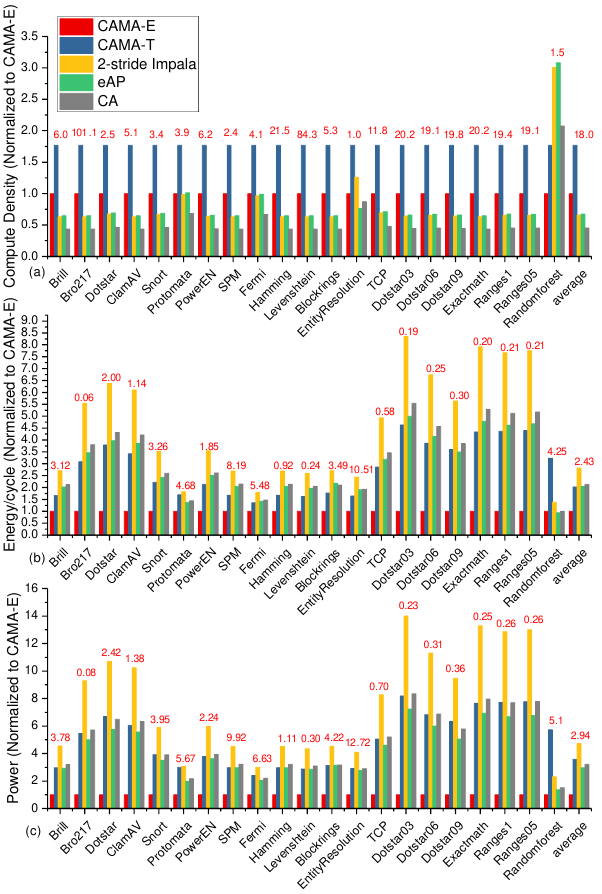}
      \vskip -2.5ex
      \caption{Comparison of the (a) compute density (b) energy per symbol, and (c) power in 28nm. All bars are normalized to the results of CAMA-E for clarity and the absolute values (in units of Gbps/$mm^{2}$, nJ/byte, and Watt) of CAMA-E are shown by red texts.}
	\label{fig:28nm}
	\vskip -3ex
\end{figure}
As mentioned in eAP and Impala, it is impossible to power-gate state matching arrays on a cycle-by-cycle bias in pipeline design as we need to know the potential next states in advance. They are calculated in the transition phase in the last cycle but overlapped with the state matching phase in one cycle in the pipeline design. Meanwhile, only active STEs will access local switch and the more active rows a switch has, the more energy it will consume. This is not included in Impala and eAP's energy model as they only consider the worst case in which all switches are accessed and the number of active rows in switch is not took into account. We update the energy/power results for CA and 2-stride Impala based on the two observations. 
\begin{figure}[t]
  \centering
      \includegraphics[scale=0.4]{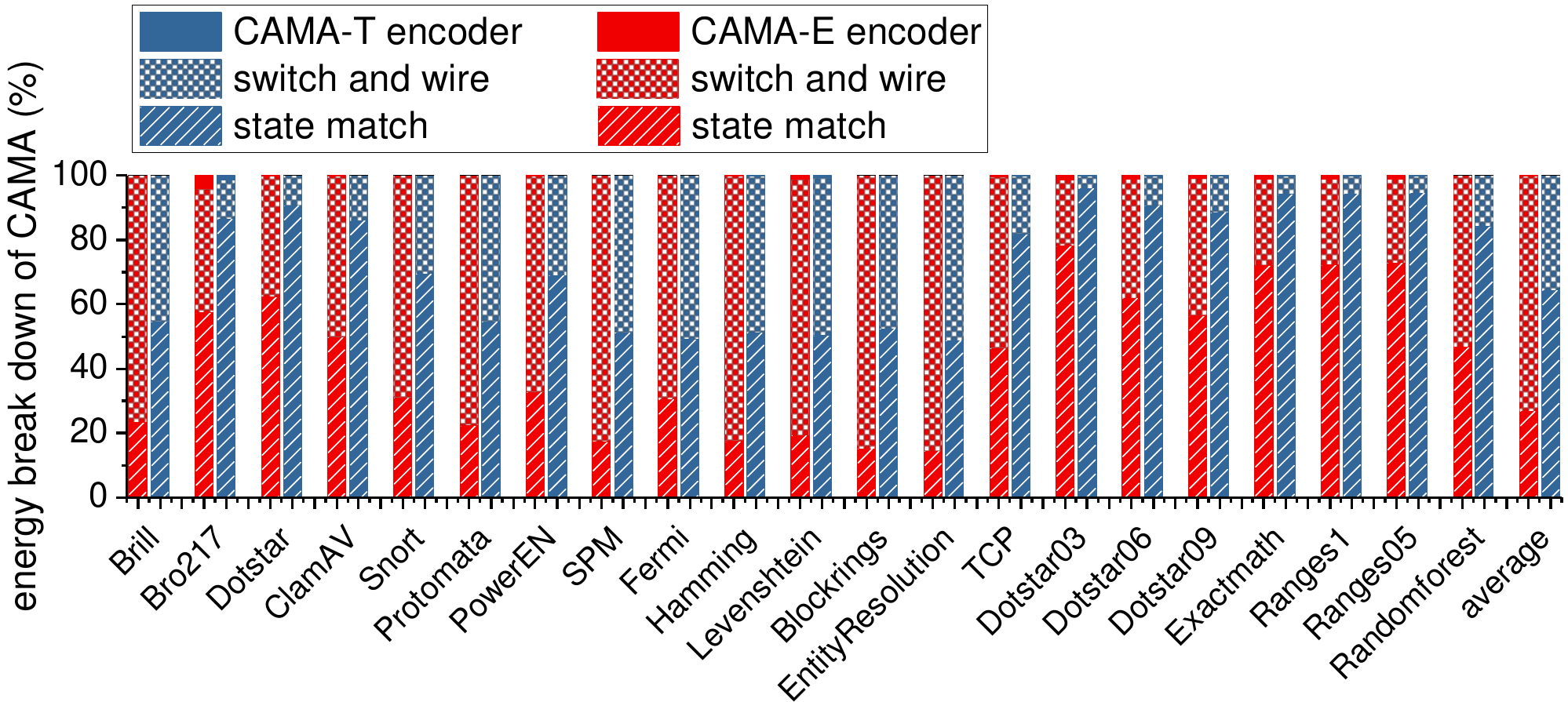}
      \vskip -2.5ex
      \caption{Energy breakdown of CAMA in different benchmarks.}
	\label{fig:break}
	\vskip -3ex
\end{figure}
As Figure~\ref{fig:28nm}(b) shows, on average, CAMA-E achieves 2.1$\times$, 2.8$\times$, 2.04$\times$ and 2.04$\times$ lower energy than CA, 2-stride Impala, eAP and CAMA-T in 28nm, respectively. This is as expected because the access energy of the state-matching CAM varies from 2.67pJ to 16.78pJ depending on the number of enabled entries which is fundamentally smaller than that of accessing all rows (16.78 pJ). In the worst benchmark (Fermi), CAME-E only consumes 7.8pJ on average due to the selectively enbled entries. CAMA-T's energy stays at 16.78pJ, which is slightly lower than CA and eAP's 19.45pJ because of the reduced CAM rows. The 2-stride Impala requires the larger energy (30.6pJ) to access two 16x256 SRAMs due to the doubled readout periphery. Particularly, both CAMA designs have higher energy on RandomForest as it works on the 32-bit mode. 
We show the energy breakdown of CAMA in Figure~\ref{fig:break}. On average, the state matching occupies 27$\%$ and 64.6$\%$, the interconnect takes 72.89\% and 35.35\% and the encoder only accounts for 0.11$\%$ and 0.05$\%$ of the total energy in CAMA-E and CAMA-T.
\par
The power consumption (see Figure~\ref{fig:28nm}(c)) further verifies the high efficiency of CAMA-E. On average, CA, 2-stride Impala, eAP and CAMA-T have 3.15$\times$, 4.71$\times$, 2.94$\times$ and 3.63$\times$ more power than CAMA-E in 28nm, respectively.
\subsection{Multi-stride Comparison}

\begin{figure}[t]
  \centering
  \vskip 0ex
      \includegraphics[scale=1.35]{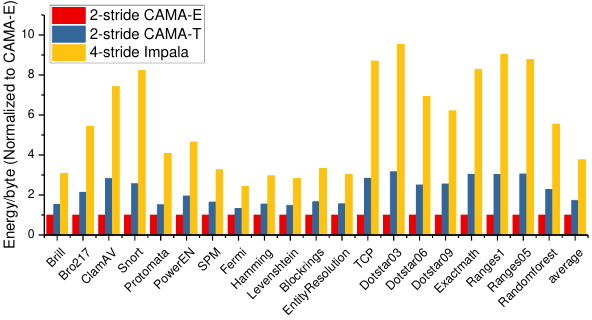}
      \vskip -2.5ex
      \caption{Energy for 2-stride CAMA and 4-stride Impala in 28nm.}
	\label{fig:stride}
	\vskip -3ex
\end{figure}

CAMA is compatible with multi-stride operation for higher throughput. Here, we compare the 2-stride CAMA with 4-stride Impala using the striding algorithm in \cite{becchi_multi_2008,becchi_improved_regular_2007}. To support the 2-stride design, CAMA's state matching array must be 64$\times$256 and the local switch should be expanded to 256$\times$256 to support a larger number of transitions and alphabet sizes. Figure~\ref{fig:stride} shows that on average 4-stride Impala consumes 2.18$\times$ and 3.77$\times$ more energy than 2-stride CAMA-T and 2-stride CAMA-E because 4-stride Impala needs four 16$\times$256 SRAM banks compared to CAMA's one 64$\times$256 bank (61.2pJ vs 22pJ).

\section{Conclusion}

This paper presents CAMA, a CAM-based in-memory architecture for efficient automata processing. CAMA co-designs hardware architecture and encoding algorithms to achieve both high compute dense and energy efficiency. It exploits an 8T 16$\times$256 CAM bank to reduce the state matching area and energy, and a 128$\times$128 RRCB as compact state transition switches. The proposed encoding optimization algorithms selects the optimal encoding to balance the compression rate and the code length for a given application, and cluster symbols into the encoding clusters for optimal CAMA performance and efficiency. 
CAMA can be designed in non-pipelined version (CAMA-E) to save energy and pipelined mode (CAMA-T) for faster speed. On average, the proposed CAMA-T shows 2.68$\times$, 3.87$\times$ and 2.62$\times$ higher compute density than 2-stride Impala, CA and eAP in 28nm while CAMA-E shows 2.1$\times$, 2.8$\times$, 2.04$\times$ and 2.04$\times$ lower energy than CA, 2-stride Impala, eAP and CAMA-T in 28nm. 